\documentclass[sigconf, authorversion,nonacm]{acmart}
\usepackage{xcolor}
\usepackage{graphicx}
\newcommand{\added}[1]{\textcolor{blue}{#1}}
\renewcommand{\added}[1]{#1}
\AtBeginDocument{%
  }

\copyrightyear{2025}
\acmYear{2025}
\setcopyright{acmlicensed}\acmConference[CHI '25]{CHI Conference on Human Factors in Computing Systems}{April 26-May 1, 2025}{Yokohama, Japan}
\acmBooktitle{CHI Conference on Human Factors in Computing Systems (CHI '25), April 26-May 1, 2025, Yokohama, Japan}
\acmDOI{10.1145/3706598.3713351}
\acmISBN{979-8-4007-1394-1/25/04}

\begin{document}
\title{Social Media for Activists: Reimagining Safety,\\ Content Presentation, and Workflows}

\author{Anna Ricarda Luther}
\email{aluther@ifib.de}
\orcid{0000-0002-1169-9297}
\affiliation{%
  \institution{Institute for Information Management Bremen GmbH}
  \city{Bremen}
  \country{Germany}
}
\affiliation{%
  \institution{University of Bremen}
  \city{Bremen}
  \country{Germany}
}

\author{Hendrik Heuer}
\email{hendrik-heuer@cais-research.de}
\affiliation{%
  \institution{Center for Advanced Internet Studies gGmbH (CAIS)}
  \city{Bochum}
  \country{Germany}}
\affiliation{
  \institution{University of Wuppertal}
  \city{Wuppertal}
  \country{Germany}
  }

\author{Stephanie Geise}
\email{sgeise@uni-bremen.de}
\affiliation{%
  \institution{Centre for Media, Communication and Information
Research (ZeMKI)}
  \city{Bremen}
  \country{Germany}
}
\affiliation{%
  \institution{University of Bremen}
  \city{Bremen}
  \country{Germany}
}

\author{Sebastian Haunss}
\email{sebastian.haunss@uni-bremen.de}
\affiliation{%
 \institution{Research Center on Inequality and
Social Policy (SOCIUM)}
 \city{Bremen}
 \country{Germany}}
 
\affiliation{%
 \institution{University of Bremen}
 \city{Bremen}
 \country{Germany}}

\author{Andreas Breiter}
\email{abreiter@ifib.de}
\affiliation{%
  \institution{Institute for Information Management Bremen GmbH}
  \city{Bremen}
  \country{Germany}}
  
\affiliation{%
  \institution{University of Bremen}
  \city{Bremen}
  \country{Germany}}

\renewcommand{\shortauthors}{Luther et al.}

\begin{abstract}
Social media is central to activists, who use it internally for coordination and externally to reach supporters and the public. To date, the HCI community has not explored activists' perspectives on future social media platforms. In interviews with 14 activists from an environmental and a queer-feminist movement in Germany, we identify activists' \added{needs} and feature requests for future social media platforms. The key finding is that on- and offline safety is their main \added{need}. Based on this, we make concrete proposals to improve safety measures. Increased control over content presentation and tools to streamline activist workflows are also central to activists. We make concrete design and research recommendations on how social media platforms and the HCI community can contribute to improved safety and content presentation, and how activists themselves can reduce their workload. 
\end{abstract}



\maketitle
\balance

\section{Introduction}
Social media has become an indispensable tool for social movements. It enables them to disseminate information rapidly, mobilize supporters across geographical boundaries, and engage with broader audiences more effectively than ever before ~\cite{Hwang2015Social, Momeni2017Social, Pandey2019A}.  Social media platforms have long been praised for facilitating political participation and amplifying the voices of marginalized communities~\cite{Hwang2015Social, Momeni2017Social, Pandey2019A}. However, social media also presents significant challenges for activists. \added{Social media platforms prioritize user engagement to collect data and generate profits, with governance shaped by corporate ownership and commercial interests~\cite{elmimouni2024shielding, Schwarz2019Facebook, Stockmann2020Media}. These priorities often conflict with activists' needs to effectively mobilize, coordinate, and disseminate information to their target audiences. Social media platforms can also empower antagonist actors such as states, corporations, or movements targeting marginalized groups~\cite{eddington2018communicative,reich2017organizational, sanches2020under}. Advancements in monitoring tools, coupled with the spread of computational propaganda tactics—such as political bots, coordinated hashtag campaigns, and manipulated viral content—along with internet shutdowns, have significantly hindered democratic participation, especially for marginalized grassroots movements~\cite{kow2020water, uldam2018social, petterson2023supporting}. These tensions motivate our research, as they emphasize the vulnerabilities activists face on social media.} Issues such as exposure to hate speech and misinformation and the growing inequalities in content visibility can \added{further} undermine the effectiveness of activism, creating hostile environments and limiting the reach of critical messages ~\cite{Castillo-Esparcia2023Evolution}.\\
\added{Research has not fully explored how activists assess these challenges and what they need to address them.} Limited research examines the \added{needs} of activists \added{relating to future} social media \added{platforms ~\cite{Castillo-Esparcia2023Evolution, elmimouni2024shielding, abokhodair2024opaque}}. Understanding these \added{needs} is critical not only to improve the effectiveness of social media as a tool for activism but also to ensure that these platforms do not inadvertently hinder the movements from driving meaningful social change. Therefore it is important to explore what activists identify as \added{needs to} work on these platforms and what features they request to \added{meet these needs}.
To this end, we conducted 14 semi-structured interviews with activists from two different social movements to answer the following research questions (RQs): What are the \added{needs of activists when using} social media for activist work~(RQ1)? What are feature requests to \added{address these needs}~(RQ2)? The two movements differ in scale, protest forms, and social media strategies. One of the movements is an environmental movement and the other is a queer-feminist movement. 

The study makes two key contributions. First, it offers insights into activists' social media \added{needs} and presents requested features. Particularly the finding that participants highlighted the critical design need for both on- and offline safety, is of high importance. The activists provided concrete suggestions for how platforms can \added{address these needs}, extending previous work primarily focused on online safety. Second, we offer research and design recommendations. Our study contributes to aligning social media design more closely with the \added{needs} of those who use it for social change: on- and offline safety, enhanced content presentation, and optimized workflows.

\section{Background}
In this section, we discuss academic work relevant to the current study. First, we examine work on the role of social media in activism and present previously identified challenges and benefits. Second, we review other efforts aimed at supporting activists relating to their social media use. 
\label{Background}

\subsection{Social Media and Activism}
Social media is an important tool for activism, influencing how movements communicate, mobilize, and coordinate their efforts~\cite{Cammaerts2015Social, gil2021whatsapp, Greijdanus2020The, Momeni2017Social}. Through social media, the dependence on traditional media outlets for information dissemination~\cite{Poell2015Social} is reduced. This change allows perspectives that are often underrepresented in mainstream media coverage~\cite{Burton2021Using, Ngidi2016Asijiki}, or suppressed by authoritarian states~\cite{Chunly2020Social} to be expressed and received by a wider audience. Social media is commonly believed to facilitate political activism by providing increased visibility for activist causes and creating low-barrier pathways for participation \cite{Greijdanus2020The, Momeni2017Social}. Social media platforms enable activists to reach supporters quickly, mobilize across geographical boundaries, and connect with audiences that share their values and beliefs~\cite{Momeni2017Social}. 
\added{Elmimouni et al. have argued, that the emphasis of social media on user engagement can function in congruence with activists' goals~\cite{elmimouni2024shielding}, when activists create content using strategies catering to algorithms (or their perception of it), leading to increased engagement, and amplifying activist voices.}
Additionally, social media supports not only external communication but also internal coordination within movements. Platforms like WhatsApp and Telegram have been widely used by activists to plan events, share logistical information, and strategize actions~\cite{gil2021whatsapp, haunss_fridays_2020, pang2020whatsapp}. Overall, research showed that social media can support activists by facilitating information dissemination and internal coordination.

Research also highlights significant challenges associated with social media use for activism. One of the most prominent concerns is the exposure to hate speech \added{directed at activists}~\cite{Castillo-Esparcia2023Evolution}. Hate speech has been defined by Zhang and Luo as any form of communication that demeans a person or a group based on race, color, ethnicity, gender, sexual orientation, nationality, religion, or political affiliation~\cite{zhang2019hate}. Hate speech on social media is a well-documented problem~\cite{castellanos2023hate, Castaño-Pulgarín2021Internet}, which disproportionately \added{targets} marginalized groups~\cite{Gelber2016Evidencing, Lingiardi2019Mapping, Miranda2023Analyzing} and is often connected to current political events~\cite{Meza2019Targets}. Activists often engage with current political events and their work frequently involves amplifying marginalized voices, making activists particularly vulnerable \added{to being targets of} hate speech. 
Another issue for activists on social media is that through algorithmic curation their content does not reliably reach their target audience~\cite{Castillo-Esparcia2023Evolution, Sannon2023Disability}. The often unequal distribution of reach can undermine the potential for easy access to widespread information dissemination. Inequalities in visibility can be partly explained by proprietary recommendation systems. While social media has reduced dependence on traditional media corporations, power has shifted to large social media corporations~\cite{Poell2015Social}. Furthermore, social media platforms have to support a variety of stakeholders with at times conflicting goals. \added{Platforms prioritize content that garners the most attention and user engagement. Activism can align with the principles of community engagement and discourse~\cite{Siddarth2020Engaging}, as for example demonstrated by the virality of environmentalist content on TikTok~\cite{wiley2021genz}. However, it does not necessarily drive high engagement in ways that are financially beneficial for platforms. These operational priorities raise the question of whether platforms can or even want to optimally support activists. Meta recently reduced the recommendation of political content on Facebook, Instagram, and Threads, showing the limited interest of platforms to function as facilitators for political actions ~\cite{meta_political_content}}. Previous work identified \added{different} social media types. \citet{koukaras2020social} have suggested that there are three types of social media networks: ``entertainment networks'', ``profiling networks'' and ``social networks''. None of these categories fully address the specific needs of social movements, which focus on internal coordination, outreach to supporters, and engagement with the broader public. This limitation affects how activists use these platforms. Activists often tailor their actions based on specific representations of what that platform `is' and how it functions~\cite{comunello2016proper}. For example, activists change their messaging strategies based on the perceived audience on the respective platforms~\cite{comunello2016proper}. Research has shown that different social media platforms enable some form of activism while hindering others~\cite{Ahuja2018The, Valenzuela2018Ties}. For example \citet{Haq2022Screenshots} highlighted that Instagram with its focus on image content provides users with few features to share or repost, limiting information propagation. Further, Twitter's ``What's Happening'' section and Instagram's Live Stream interface influence coalition-building efforts~\cite{Jones2021Online}. \added{TikTok functionalities have been identified as fostering the creation of creative political content~\cite{abbas2022tiktok}.
Activists also tailor their actions to their understanding of social media logics (as conceptualized for example by ~\cite{van2013understanding}). This has led to the prioritization of virality and other short-term goals over long-term goals for the use of social media for activism~\cite{Siddarth2020Engaging}. However, Hutchinson et al. argue that activists do not sufficiently leverage these principles for their own goals such as increased visibility across social media spaces~\cite{hutchinson2021micro}.} 
Thus, activists face challenges on social media, including hate speech, inconsistent reach, and navigating platform-specific functionalities and audiences.

\subsection{Previous Research on Supporting Activists on Social Media}
Previous research has explored various strategies to support activists \added{to navigate existing challenges with social media}, focusing on areas like hashtag usage~\cite{Bisafar2020act, Karbasian2021Impro, Malik2018How} and messaging strategies~\cite{Siddarth2020Engaging}, and the design of social media interfaces~\cite{Haq2022Screenshots, Jones2021Online}. 

One area of support involves teaching activists to use social media more strategically. \citet{Bisafar2020act} showed activists how to identify the identities of their followers through the use of hashtags. This knowledge empowered young activists to refine their activism efforts by tailoring messages more effectively to their audience. Specifically, analyzing patterns of co-occurring hashtags, relevant topics, and sentiments has been shown to enhance these efforts by boosting mobilization~\cite{Karbasian2021Impro}. In campaigns promoting diversity in engineering, the strategic use of hashtags significantly increased reach and engagement~\cite{Karbasian2021Impro}. Multivocality, which refers to the ability of users to participate through multiple modes, messaging platforms, and actors is a key factor in the success of hashtag activism campaigns, such as the \#ILookLikeAnEngineer movement~\cite{Malik2018How}. Additionally, real-time insights into user participation have been achieved through machine learning systems that classify user types engaging in hashtag activism~\cite{Karbasian2018Real}. By using strategic social media practices, activists can significantly increase the reach and effectiveness of cause-driven campaigns. \\
Studies have also researched social media practices of different activist groups~\cite{Siddarth2020Engaging, LeFevre2023Abortion, Ismail2019Empowerment}, \added{providing insights into how different activists experience these platforms}. By interviewing and observing activists, researchers have gained an understanding of how messaging strategies are self-assessed by the activists, \added{showing the conflicts between short-term and long-term movement goals exacerbated by social media}~\cite{Siddarth2020Engaging}. \added{For pro-choice activists, the research identified a lack of community support as a major challenge, prompting design concepts focused on fostering support networks}~\cite{LeFevre2023Abortion}. \added{Further, the experience of health workers on social media has been investigated, showing that community health workers use social media to challenge gender norms, become digital citizens, and provide better quality healthcare}~\cite{Ismail2019Empowerment}. Based on these insights, previous research identified opportunities for social media to support these efforts~\cite{Siddarth2020Engaging, Ismail2019Empowerment} \added{and offers a design probe and a prototype to better support their activism~\cite{LeFevre2023Abortion, tadic2023design}. Examining these challenges and solutions offers insights into activists' struggles with platform limitations and serves as a valuable reference point for our work.} 
 
\added{These studies highlight specific challenges faced by activist groups and introduce design solutions to support their efforts. However, while previous work has drawn on observed experiences, our research adds the perspective of directly engaging activists of two distinct social movements to identify their needs for a future social media platform and the features they believe are essential to address these needs.}

\section{Methods}
We employed a qualitative approach consisting of semi-structured interviews and a card-sorting task. Semi-structured interviews have been used in previous studies on social media practices~\cite{Islam2020Deep, LeFevre2023Abortion, Siddarth2020Engaging} and eliciting future visions across various domains ~\cite{Alsayed2024Exploring, Guo2024Designing, Mlynar2022AI}. To complement the interviews, a card-sorting task was conducted to reveal how activists prioritize their social media \added{needs}. This technique, effective both online and offline ~\cite{Bussolon2006Online}, uncovers key priorities. By combining semi-structured interviews and a card-sorting task, this research provides valuable insights into activists' \added{needs} and desired features for future social media platforms. 

\subsection{Participants}
The primary objective of this study was to explore the \added{needs} of different social movements associated with the diverse ways in which social media is employed for activism. Accordingly, we decided to recruit participants from two different social movements: one is part of the environmental movement and the other is part of the queer-feminist movement. \added{We selected the environmental movement, as it was one of the largest protest mobilizations in Germany in the last years (omitted for reasons of anonymization), holding great societal relevance. The local chapters organize the environmental movement as part of a nationwide and global network. This movement is a specific representative of de-centralized social movements, allowing for insights into the use of social media both for internal and external communications of a de-centralized environmental movement. To understand what needs are specific to an individual movement and what needs are more broadly applicable, we selected a second movement as a contrasting case. The selection criteria for this are} notable differences in size, predominant protest form, and social media strategies. \added{Based on these criteria, we chose to recruit participants from a queer-feminist movement.}

The environmental movement is of a larger scale than the queer-feminist movement, operates through well-established nationwide and global networks, \added{and focuses organized campaigns and specifically widespread public demonstrations.} Social media platforms are predominantly used as broadcast channels to disseminate information about their topic of activism and about upcoming activist actions\added{, most commonly demonstrations} (omitted for reasons of anonymization). Also, messenger apps are used for internal coordination as well as reaching supporters (omitted for reasons of anonymization). Despite the national and global networks, the local chapters are active in very localized contexts. In contrast, the queer-feminist movement is smaller than the environmental movement (with a total of 70 members nationwide) and more grassroots-oriented, with activist actions occurring in more localized contexts (omitted for reasons of anonymization). These smaller local chapters are a part of a nationwide and global network. The predominant form of activism is visualizing incidents of street harassment at the place where they occurred, taking pictures of them, and then uploading them on Instagram. The incidents of street harassment are anonymously reported to the activists via Instagram. For internal coordination, messenger services are also commonly used. \added{For the environmental movement, social media is used to reach people and mobilize in the offline sphere. For the queer-feminist movement social media is part of their form of protest by linking the on- and offline sphere.} Investigating these different movements provides a basis for examining the varied ways in which social media is utilized within activism and what \added{needs} are associated with it, from large-scale mobilization efforts to more community-based advocacy.

We recruited a total of 14 participants (nine female), with an average age of 24.14 years (SD = 2.14, min = 19, max = 27). Nine participants are active in an environmental movement (five female), and five participants are part of a queer-feminist movement (five female). The demographic characteristics of the participants from the environmental movement are comparable to the age and gender distribution of the respective movement, as reported by previous research (omitted for reasons of anonymization). There is no prior research on the demographic distribution for the German local chapters of the queer-feminist movement. However, the board of the nationwide umbrella organization, which collected data on the age distribution and pronouns (which are not classified according to gender categories of male, female, or diverse) of its members across Europe, has stated that the mean age is 25, with the minimum age of 15 and the maximum age of 47. No statements relating to the standard deviation of the age distribution were made.  

To recruit participants, the official communication channels of each movement were utilized. In the case of the environmental movement, participants were recruited by emailing all local chapters through the contact information published on the official website. As there is no such overview of local chapters and their contact information for the queer-feminist movement, the nationwide umbrella organization of the movement was contacted via email, and they forwarded the request to all local chapters via email and the respective WhatsApp groups.

Recruitment was a time-consuming endeavor. Reliance on umbrella organizations introduced extra layers of communication, which slowed down responses. Tailoring outreach to decentralized or less formally structured activist networks may prove helpful for future recruitment efforts. We recommend exploring more targeted recruitment approaches, such as using social media platforms frequented by activists or engaging directly with grassroots organizers. Each participant received 30€ (\textasciitilde\$33.3) for their participation.

\subsection{Procedure}
We conducted 14 semi-structured interviews with a card-sorting task. The participants were located across various regions of Germany. Thus, 12 out of the 14 interviews were conducted using videoconferencing tools, while two were conducted in person. All interviews were audio-recorded and transcribed by a professional transcription service. The study received IRB-equivalent approval. 
\begin{figure*}
    \Description{The figure shows which tasks our participants listed as part of their activist work and which functionalities of Instagram and Messengers are associated with each task. The activist tasks presented on the left-hand side are separated into External Communication and Internal Communication. External communication is further divided into Networking, Disseminate Information, Activist Actions, Educational Work, and Coordination. Internal Communication is further divided into Task Allocation, Material Creation, Information Exchange, Material Procurement, and Teambuilding. On the right-hand side, the functionalities of Instagram are listed: Content Creation, Engagement \& Interaction, Discovery \& Browsing, Account Management, and Content Moderation. Below, the functionalities of Messengers are listed: Group Chats, Individual Chats, Channels, Privacy \& Security, and Collaborative Tools. Each functionality is then linked to an activist task, given that at least two participants reported using this functionality for the respective task. Content Creation is linked to Networking, Disseminate Information, Activist Actions, and Educational Work. Engagement and Interaction are linked to Networking, Disseminate Information, Educational Work, and Coordination. Discovery \& Browsing is linked to Networking and Educational Work. Account Management is linked to Networking and Coordination. Content Moderation is linked to Disseminate Information and Educational Work. For the Messenger functionalities Group Chats are linked to Networking, Task Allocation, Material Creation, Information Exchange, and Teambuilding. Individual Chats are linked to Networking, Task Allocation, Information Exchange, Material Procurement, and Teambuilding. Channels are linked to Disseminate Information, Information Exchange, and Material Procurement. Privacy \& Security is linked to Task Allocation and Information Exchange. Collaborative Tools are linked to Material Creation and Teambuilding.}
    \caption{This figure gives an overview of activist tasks and the role of social media use for these tasks. Each link from the functionalities of Instagram and messenger services (WhatsApp, Telegram, and Signal) to the categories of activist tasks indicates that at least two participants mentioned the respective functionality as important for the respective activist task.}
    \label{fig:ActivistTasks}
    \centering
    \includegraphics[width=0.8\textwidth]{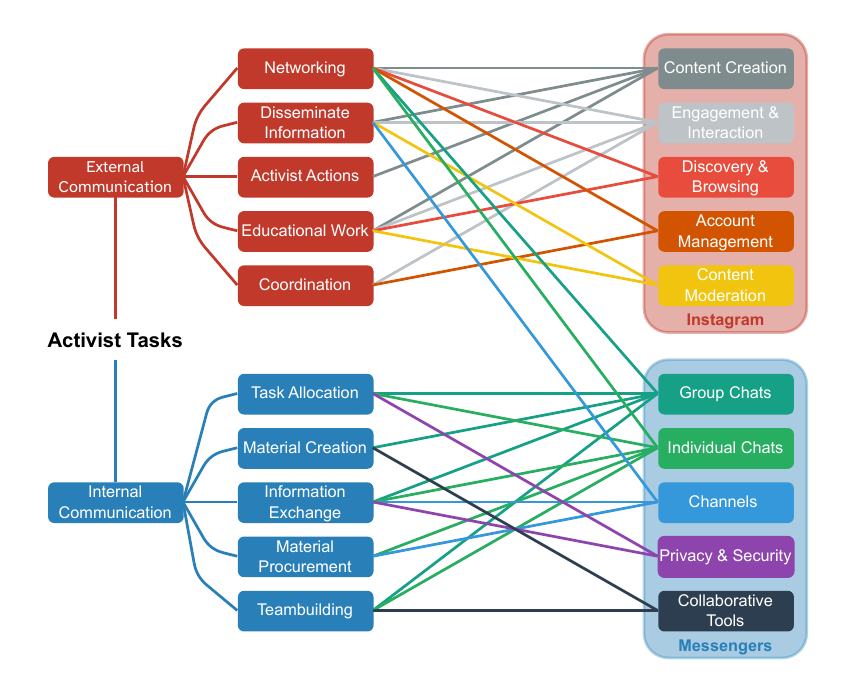}
\end{figure*}
In the introductory phase of the interview, each participant gave informed consent in line with GDPR and answered demographic questions. Then, each participant was asked about their everyday activist work and the role of social media within. We first asked the participants about the most recent activist events and the everyday activist work, to get an overview of their activist work. All tasks that fall within their area of responsibility were collected. For each task, we asked if they used social media to fulfill the task and if yes, which social media platforms they used in which way \added{(Figure \ref{fig:ActivistTasks} gives an overview of our categorization of the activist tasks and their connection to social media. A description of the suggested categories can be found in the Appendix \ref{ActivistTasks}).} 

During the card-sorting task, participants were asked to categorize the collected tasks according to two criteria: first, whether a task was performed with or without technical support (X-axis); and second, how they assessed the impact of the task for their activism (Y-axis). Participants were encouraged to provide a rationale for their answers and to suggest potential improvements to the process.
Subsequently, the role of social media in activism and potential improvements were discussed. The participants were requested to indicate which tasks they considered taking up most of their time, which tasks were most laborious, and which were most frustrating. The participants were encouraged to elaborate on their reasoning. Furthermore, each participant was asked how they assessed their success in achieving their objectives. Then, participants were asked what they perceived as the advantages and disadvantages of social media for their activism. 
Participants were asked about desired improvements to current social media functionalities. Additionally, they were tasked with imagining their ``perfect social media platform'' for their activism. \added{This use of imagination draws on theoretical work highlighting how future-oriented discourse fosters shared sociotechnical imaginaries~\cite{natale2020imagining, messeri2015greatest}. By envisioning ideal platforms, participants could articulate needs beyond current limitations, offering insights into transformative possibilities for social media design.}

\subsection{Analysis}
We analyzed the transcribed material of the semi-structured interviews using thematic analysis~\cite{braun_using_2006}. The first author read the texts multiple times and moved back and forth through the material. An inductive open coding of the material was conducted, following axial coding principles~\cite{corbin2014basics}. The codes were discussed with the second author in regular meetings until a unanimous agreement was reached. Following the clustering, splitting, and merging of various codes to form cohesive groups, we discerned categories and subcategories, which were then clustered into themes, serving as high-level \added{needs. While our research aim was to identify activist needs, our participants repeatedly reported problems when using social media for activism. Thus, several identified needs address changes necessary to solve the mentioned problems.} Only statements \added{explicitly focused on} activist social media use served as a basis for analysis\added{; surrounding or contextual statements related to non-activist social media use were not included. We asked the participants 
to report on their experiences on any social media platform used for activism. The activists predominantly described their experiences on Instagram.}

\section{Results}
This section presents the findings on the \added{needs} associated with social media use for activism (RQ1) and the desired features to \added{meet these needs} (RQ2). We categorized these \added{needs} into three high-level \added{needs}: on- and offline safety, enhanced content presentation, and optimized workflows. The \added{needs} and feature requests, organized by their frequency, are summarized in Table~\ref{tab:table1} \added{and an overview over which activists mentioned which need can be found in Table \ref{tab: table2}}. \added{All listed feature requests were explicitly requested by the participants and not inferred during analysis}. Activists from the environmental movement are identified with an `E', followed by their participant number (e.g., E07), while activists from the queer-feminist movement are identified with a `Q' and their participant number (e.g., Q03). If the participants mentioned social media platforms, they referred to Instagram, unless mentioned otherwise.

\begin{table*}
\caption{The \added{needs} associated with social media use for activism (RQ1) and the requested features to \added{meet these needs} (RQ2) as identified through semi-structured interviews with a card-sorting task with 14 activists from two social movements (an environmental and a queer-feminist movement).
}
\label{tab:table1}
\Description{This table presents high-level needs, needs, and feature requests related to using social media for activism. The high-level needs column lists 3 high-level needs: On- and Offline Safety; Enhanced Content presentation; and Optimized Workflows. The Needs column lists seven needs that relate to high-level needs: Physical Safety; Reducing Exposure to Hate Speech; Reaching Target Audience; Controlling Scrolling Behavior; Managing Misinformation; Reducing Workload, Increasing Customizability. The Feature Requests column lists 29 feature requests corresponding to the Needs: Non-public comments; Self-deleting messages; More levels of anonymity; Account creation without phone number; Link between accounts and user identities; Reliable filtering of hate speech; Educational explanations of filtering; Increased consequences for hate speech; Custom recommendation settings; Insights into recommendation system; Chronological feed; No reduction of political content; Enhanced profile organization; Disable infinite scrolling; Toggle infinite scroll feature; Intervention after set scrolling duration; Custom home screen; Removal of short videos; Flag misinformation with labels; Reliable filtering of misinformation; Unified platform for all use cases; Integrated planning module; Expanded reaction options; Automatized text editing; Extended image editing options; Extended video editing options; Customizable display options; Customizable screen-reading; Customizable distance between posts}
    \begin{tabular}{l l l}
    \toprule
    \textbf{High-level Needs} & \textbf{Needs} & \textbf{Feature Requests} \\
    \midrule
    \textbf{On- and Offline Safety} & Physical Safety & Non-public comments\\
    &\added{(100\% Queer-feminists; 55.6\% Environmentalists)}& Self-deleting messages\\
    && More levels of anonymity \\
    &&Account creation without phone number\\
    && Link between accounts and user identities\\
    & Reducing Exposure to Hate Speech & Reliable filtering of hate speech \\
    &\added{(100\% Queer-feminists; 55.6\% Environmentalists)}& Educational explanations of filtering\\
    && Increased consequences for hate speech \\    
    \hline
    \textbf{Enhanced Content Presentation} & Reaching Target Audience & Custom recommendation settings\\
    &\added{(100\% Queer-feminists; 100\% Environmentalists)}& Insights into recommendation system\\
    && Chronological feed\\
    && No reduction of political content\\
    && Enhanced profile organization \\
    & Controlling Scrolling Behavior & Disable infinite scrolling\\
    &\added{(40\% Queer-feminists; 55.6\% Environmentalists)}& Toggle infinite scroll feature \\   
    && Intervention after set scrolling duration\\
    && Custom home screen\\
    && Removal of short videos \\
    & Managing Misinformation & Flag misinformation with labels\\
    &\added{(80\% Queer-feminists; 44.4\% Environmentalists)}& Reliable filtering of misinformation \\
    \hline
    \textbf{Optimized Workflows} & Reducing Workload & Unified platform for all use cases\\
    &\added{(60\% Queer-feminists; 88.9\% Environmentalists)}& Integrated planning module \\
    && Expanded reaction options \\
    && Automatized text editing \\
    && Extended image editing options \\ 
    && Extended video editing options\\
    & Increasing Customizability  & Customizable display options \\
    &\added{(60\% Queer-feminists; 55.6\% Environmentalists)}& Customizable screen-reading \\
    && Customizable distance between posts \\
    \bottomrule
\end{tabular}
\end{table*}

\subsection{Needs of Activists for Social Media Use~(RQ1)}
\subsubsection{On- and Offline Safety} 
Participants highlighted on- and offline safety as \added{an important need}. They reported online threats and in-person assaults, which made them feel unsafe during activist actions. Especially a combination of online visibility and real-life activism heightened their sense of vulnerability, leading some to reduce their social media presence~(Q01, Q05, Q06, E07). Hate speech on social media \added{directed at the activists} was another significant issue, contributing to mental exhaustion and distorting perceptions of public opinion. The following subsections explore the \added{needs} relating to on- and offline safety.
\begin{table*}
\caption{An overview of which participants mentioned which needs when using social media for activism (RQ1) in semi-structured interviews with a card-sorting task with 14 activists from two social movements (an environmental and a queer-feminist movement).
}
\Description{This table presents an overview of which participants mentioned which need relating to using social media for activism. The table has 15 columns, the first column is the needs column and the following 14 columns represent each participant, chronologically ordered based on their numbering. The needs column lists seven needs: Physical Safety; Reducing Exposure to Hate Speech; Reaching Target Audience; Controlling Scrolling Behavior; Managing Misinformation; Reducing Workload; Increasing Customizability. The following columns show the participant numbers as header and a checkmark for each need that was named by the respective participant. The need Physical Safety was mentioned by the following participants: Q01, E02, Q03, Q04, Q05, Q06, E07, E08, E12, E13. The need Reducing Exposure to Hate Speech was mentioned by the following participants: Q01, E02, Q03, Q04, Q05, Q06, E07, E11, E12, E13. The need Reaching Target Audience was mentioned by the following participants: Q01, E02, Q03, Q04, Q05, Q06, E07, E08, E09, E10, E11, E12, E13. The need Controlling Scrolling Behavior was mentioned by the following participants: Q04, Q05, E08, E09, E10, E12, E13. The need Managing Misinformation was mentioned by the following participants: Q01, Q04, Q05, Q06, E10, E11, E12, E13. The need Reducing Workload was mentioned by the following participants: E02, Q04, Q05, Q06, E07, E08, E09, E10, E11, E12, E13. The need Increasing Customizability was mentioned by the following participants: Q01, E02, Q04, Q05, E07, E09, E10, E12.}
\label{tab: table2}
\resizebox{\textwidth}{!}{%
    \begin{tabular}{l l l l l l l l l l l l l l l}
    \toprule
    \textbf{Needs} & Q01 & E02  & Q03  & Q04  & Q05  & Q06  & E07 & E08 & E09 & E10 & E11 & E12 & E13 & E14  \\
    \midrule
    Physical Safety &$\checkmark$&$\checkmark$&$\checkmark$&$\checkmark$&$\checkmark$&$\checkmark$&$\checkmark$&$\checkmark$&&&&$\checkmark$&$\checkmark$& \\
    Reducing Exposure to Hate Speech &$\checkmark$&$\checkmark$&$\checkmark$&$\checkmark$&$\checkmark$&$\checkmark$&$\checkmark$&&&&$\checkmark$&$\checkmark$&$\checkmark$& \\
    Reaching Target Audience &$\checkmark$&$\checkmark$&$\checkmark$&$\checkmark$&$\checkmark$&$\checkmark$&$\checkmark$&$\checkmark$&$\checkmark$&$\checkmark$&$\checkmark$&$\checkmark$&$\checkmark$&$\checkmark$\\
    Controlling Scrolling Behavior & & & & $\checkmark$&$\checkmark$ & & &$\checkmark$ &$\checkmark$ &$\checkmark$ & &$\checkmark$ &$\checkmark$ &\\
    Managing Misinformation &$\checkmark$&&&$\checkmark$&$\checkmark$&$\checkmark$&&&&$\checkmark$&$\checkmark$&$\checkmark$&$\checkmark$&\\
    Reducing Workload & &$\checkmark$&&$\checkmark$&$\checkmark$&$\checkmark$&$\checkmark$&$\checkmark$&$\checkmark$&$\checkmark$&$\checkmark$&$\checkmark$&$\checkmark$&\\
    Increasing Customizability &$\checkmark$&$\checkmark$&&$\checkmark$&$\checkmark$&&$\checkmark$&&$\checkmark$&$\checkmark$&&$\checkmark$&&\\
    \bottomrule
    \end{tabular}%
    }
\end{table*}

\added{\textbf{Physical Safety:}}
\added{Activists expressed the need} to be and to feel safe, especially physically safe~(Q01, E02, Q03, Q04, Q05, Q06, E07, E08, E12, E13), as they currently perceive a lack of safety~(Q01, E02, Q03, Q04, Q05, Q06, E07). Ten participants reported online assaults (Q01, E02, Q03, Q04, Q05, Q06, E07, E11, E12, E13). Q05 described: ``\textit{In addition to rape wishes, violence wishes, you also read death threats}''. All members of the queer-feminist movement were either personally confronted with or witnessed in-person assaults. This created a persistent fear of physical harm and a strong sense of insecurity during activist actions~(Q01, E02, Q03, Q04, Q05, Q06, E07). Q06 feared: \textit{"I also see a danger that it [the negative comments they get during activist actions] could become physical"}. As the majority of actions described by activists take place at least in part in person, this is especially problematic. Participants perceived the combination of both being present on social media and performing activist actions mostly in person as unsafe. As a result, several participants~(Q01, Q05, Q06, E07) avoid posting content on their activist group’s social media account where they could be identified: 
\begin{quote}
    ``\textit{I definitely wouldn’t just post a video of myself in the story [...] Because then I have the feeling that [local city] isn’t that big and people could find me.}'' (Q01)
\end{quote}

\added{\textbf{Reducing Exposure to Hate Speech:}}
Participants also addressed the issue \added{that they commonly receive hate comments directed at them}. \added{This} had a significant impact on activists' experiences on social media. Ten participants reported exposure to hate speech, which led to mental exhaustion or frustration~(Q01, E02, Q03, Q04, Q05, Q06, E07, E11, E12, E13). Q01 complained: ``\textit{If you are simply insulted for no reason on social media, that’s very frustrating}''. \added{Frequently viewing} hate speech \added{comments, also if not directed at them personally,} affected activists' perceptions of the broader discourse on their issues. As E10 noted, this often leads to a distorted view of public attitudes: 
\begin{quote}
    ``\textit{I project it too much onto real life. So I think, just because Rainer writes on Facebook that all women are shit, I have the feeling that, okay, men think all women are shit.}''
\end{quote} 
Participants expressed reducing exposure to hate speech on social media as an important \added{need}: ``\textit{I don't want the hate}''~(Q01). Similarly, E11 wished: ``\textit{If it were ideal, then, of course, it wouldn't happen that people could be bullied online}‘‘ and E12 expressed that the ideal social media platform would be a safe space where users do not have to fear hate or incitement. Activists highlighted reducing exposure to hate speech as a key priority to foster a more supportive environment for activism.

\subsubsection{Enhanced Content Presentation}
Participants identified enhanced content presentation as an important \added{need} when using social media for activism. Issues such as unreliable reach on social media, the negative effects of infinite scrolling \added{(infinite scrolling refers to a design feature that automatically loads more content as a user scrolls down a page, resulting in a seemingly endless content stream~\cite{Rixen2023The})}, and the spread of misinformation significantly hindered their efforts. The following subsections explore the \added{needs} associated with enhanced content presentation.

\added{\textbf{Reaching Target Audience:}}
All participants reported that reaching their target audience with their content is an integral part of activist work. All participants view the possibility of social media to gain reach for their content as a key advantage. Several issues hinder the participants from effectively doing so. First, participants criticized the unreliable reach provided by Instagram~(Q01, Q04, Q05, E07, E08, E09, E11). This unreliable reach was assessed as disadvantageous~(Q01, Q04, Q05, E07, E08, E09, E11) and elicited negative emotions such as anger, frustration, and annoyance~(Q01, Q04, E08, E11). 

Q04 said: 
\begin{quote}
``\textit{It’s simply frustrating to have the feeling that so and so many people are following us now, and the post has somehow been played out to only a third of these people [on Instagram].}'' 
\end{quote}

As a potential reason for this unreliable reach, participants listed that the curation system is ``completely intransparent''~(E08) and that they have ``zero influence on it~''~(Q06). Thus, the participants express significant frustration with Instagram’s unreliable reach, which they see as a major obstacle \added{to address their need to effectively engage their target audience.}

\added{\textbf{Controlling Scrolling Behavior:}}
Managing their scrolling behavior is another key concern for the activists, as it significantly impacts their ability to stay focused and engaged in their work. This focuses on being able to influence how the participants scroll through their feeds. Participants assessed the infinite scroll as negatively impacting activist work~(Q04, Q05, E08, E09, E10, E12, E13), as it leads to \textit{doomscrolling} and hinders participants from actually doing activist-related tasks: 

\begin{quote}
    ``\textit{You can also be [...] doomscrolling, where you watch one bad climate news story after another [...] But at some point you just watch these videos and don’t get active.}''~(E12)
\end{quote}

\added{\textbf{Managing Misinformation:}}
Participants consistently expressed the desire for better mechanisms to manage misinformation~(Q01, Q04, Q05, Q06, E10, E11, E12, E13). Spreading accurate educational information is central to their activist work. The presence of misinformation not only distorts public understanding but also undermines their educational efforts. This issue underscores the urgent need for more effective tools and strategies to combat misinformation on social media.

\subsubsection{Optimized Workflows}
Participants highlighted two key \added{needs} for improving their social media workflows: reducing workload and increasing customizability.
First, the activists reported limited time resources, which affects their ability to manage social media effectively alongside other responsibilities. Second, they expressed a strong desire for increased customizability to better tailor their content and to better engage diverse audiences. The following subsections will explore these \added{needs} in detail.

\added{\textbf{Reducing Workload:}}
A clear majority of participants identified a lack of time resources for activism-related purposes, both for themselves and for their group members, as a challenge~(E02, Q04, Q05, Q06, E07, E08, E09, E10, E11, E12, E13). The time it takes to manage social media, organize events, and respond to online interactions next to their jobs or studies often exceeds the available time resources~(E02, Q04, Q05, Q06, E07, E08, E09, E10, E11, E12, E13). As a result, reducing the workload associated with their activist efforts is viewed as \added{a critical need}. 

\added{\textbf{Increasing Customizability:}}
Participants described increased customizability \added{as important} (Q01, E02, Q04, Q05, E07, E09, E10, E12). Q04 said: ``\textit{Being able to customize [the display options of Instagram] is really valuable}''. Customizability is also deemed important to be more inclusive to diverse audiences~(Q01, E02, Q04, Q05, E07, E09, E10, E12). Q04 elaborated on the issue of providing accessible content for every potential recipient on Instagram:
\begin{quote}
``\textit{No matter how hard you try, you’ll never be able to reach everyone […] For neuro-diverse people, it might be cool, if it’s not black and white, but black and a beige tone [...] Another person from the Germany-wide activism network, who has MS [multiple sclerosis], for her, the contrast should be as small as possible [...] I felt so overwhelmed by all these people receiving it [the Instagram post] and wanting to do it justice that I just let it go.}''
\end{quote}
This shows that the options on social media for customization are perceived as not sufficient and that this influences activists' workflows. 

\subsection{Feature Requests of Activists for Social Media Platforms~(RQ2)}

\subsubsection{On- and Offline Safety}
Participants emphasized the need for specific features to improve safety on social media. Key requests included enhanced anonymity options to protect physical safety and better tools to reduce exposure to hate speech. The following subsections present these feature requests and their implications for creating a safer environment for activists.

\added{\textbf{Physical Safety:}}
Participants requested several features that aim to increase safety. The majority of requested features relate to anonymity on social media platforms~(Q01, E02, Q03, Q04, E07, E08, E12, E13, E14). Enhancing anonymity options is often mentioned as an effective way to improve the safety of social movement members~(Q01, Q05, Q06, E07, E11, E13). To increase anonymity, participants requested non-public comments~(E12), self-deleting messages~(E02, E11), and more levels of anonymity~(Q05, E07, E12, E13) based on self-defined group association. For messenger services, in this case, Signal and WhatsApp, five participants~(E02, Q04, Q06, E08, E11) requested that their account is not tied to their phone number. Relating to the users from which the participants fear online assaults, anonymity is also partly assessed as a reason for such behavior~(Q01, Q03, Q04, E12, E13). Some participants requested that a social media account should be tied to the personal identity of the user to ensure accountability~(Q01, Q03, E12): 
\begin{quote}
    ``\textit{The platform should be integrated around your own personal identity, that you can’t create [a] thousand fake accounts, and when one is blocked, I continue to insult using the others.}'' (E12)
\end{quote}

\added{\textbf{Reducing Exposure to Hate Speech:}}
Participants requested various features to reduce the exposure to hate speech. Several participants requested reliable filters for hate speech~(Q01, Q04, Q05, Q06, E10, E12). 

Q05 described: 
\begin{quote}
    ``\textit{If somebody drops the N-word, you don't have to delete it yourself. Instead, it is automatically not posted at all.}''
\end{quote}
For the queer-feminist movement, Q01 highlighted the specific need for tailored filtering mechanisms related to their activism. Q01 experienced that reports of street harassment --- an integral part of their activism --- are filtered out when using currently available filters. Therefore, she requested that hate speech detection should be able to distinguish between the reports of sexual harassment and hate speech directed at her. Participants also suggested educational efforts coupled with more reliable filters~(Q05, Q06, E10). Q05 suggested, that platforms should display a message explaining why the content is assessed as hate speech and removed. Lastly, more consequences for users spreading hate speech~(Q03, Q06, E12, E13) were requested. E12 referred to increased legal consequences, while Q06 and E13 called for a strike system, where users would receive warnings for violations, and after three offenses, their accounts would be permanently deleted.

\subsubsection{Enhanced Content Presentation}
Participants identified enhanced content presentation as a key to improving social media use in activism. This includes several feature requests aimed at increasing agency and transparency in content curation, providing greater control over scrolling behavior, and improving the management of misinformation. The following subsections will explore these feature requests in detail.

\added{\textbf{Reaching Target Audience:}}
Several participants requested insights into the algorithmic curation system of Instagram~(Q06, E08, E10, E12, E14). These insights should explain how the curation system works and why what content is displayed to their followers. The activists expected that they could use this knowledge to better reach their target audience. Some participants requested co-creation opportunities concerning the algorithmic curation system~(Q06, E08, E10, E12, E14). Other participants, however, wished for no algorithmic curation at all and a return to a chronological feed~(Q01, Q04, E09, E13). 

During the time of our interviews, Instagram announced the plan to reduce political content on the platform~\cite{noauthor_update_nodate}. Some participants~(Q01, E08, E14) specifically demanded no reduction of political content. The option to have a more detailed and well-structured profile~(E07, E13) was also requested to better reach their target audience with their content.

\added{\textbf{Controlling Scrolling Behavior:}}
Several feature requests were formulated to \added{meet the need} of increased control over the scrolling behavior~(Q04, E08, E09, E10, E12, E13). Some participants requested the replacement of the infinite scrolling feature or the possibility of turning the infinite scroll off~(E12). Another feature to increase control over scrolling behavior is a digital nudge after some time of scrolling: 
\begin{quote}
    ``\textit{When you've been scrolling for a while, [the platform] asks: ``Hey, are you okay? Are you doing what you really want to do?''}''~(E12)
\end{quote} 
Q04, E08, E09, and E10 suggested the complete removal of short videos from the platform as they assessed them as a key factor in promoting infinite scrolling~(Q04, E08, E09, E10). Lastly, the introduction of a home screen was suggested, such that content is not immediately visible when the app is opened~(E12).

\added{\textbf{Managing Misinformation:}}
Participants differed in their feature requests relating to managing misinformation. Q01, Q04, E11, and E12 requested reliable filtering of misinformation, such that it is removed before posting. Q05, Q06, Q07, E08, and E10 preferred better labeling of misinformation. Q06 and E10 further requested that this labeling should be coupled with the display of third-party sources explaining why it is misinformation.

Q06 criticized Instagram for lacking fact-checking: 
\begin{quote}
    ``\textit{On Twitter, for example, there you can see: ``This video shows false facts. That's how it actually is''. And that's a control instance you don't have at all on Instagram.}''
\end{quote} 
This statement is surprising, considering that Instagram uses external fact-checkers and labels misinformation. This shows that existing efforts were either not noticed or not perceived as sufficient.

\subsubsection{Optimized Workflows}
Participants highlighted the need for optimized workflows. This includes requests for tools that streamline tasks, automate repetitive processes, and support collaboration across platforms. Additionally, participants emphasized the importance of customizability, particularly in terms of display and interaction options, to better meet their needs and those of their audiences. The following subsections will describe the features requested to \added{address these needs}.

\added{\textbf{Reducing Workload:}}
Participants emphasized the need for tools and strategies that streamline tasks, and automate repetitive processes, enabling them to focus their limited time on high-impact activities. The participants requested a wide range of features aiming to reduce workload. Several participants wished for a unified, multiprotocol platform that would streamline their efforts across multiple social media channels. Such a platform would allow them to manage content and interactions on all platforms from a single interface, reducing the workload of posting and monitoring updates separately. Additionally, it would facilitate collaboration among activists by centralizing communication and resource sharing, making coordinated efforts more efficient and less time-consuming~(E02, Q05, E09). Further, more nuanced reaction tools were requested~(Q03, Q05):
\begin{quote}
    ``\textit{It would be cool if every app had something like reactions. In WhatsApp and Instagram, you can just like [a message] and then the heart appears. But with Discord, I think it's cooler that you can use different emojis.}''~(Q03)
\end{quote}

The activists emphasized the importance of integrating such tools directly into the social media platforms they use. They noticed higher participation rates when engagement options were directly accessible from the respective interface. They found that making participation just a click away, for example through streamlined sharing or commenting features, significantly lowered barriers and encouraged more people to get involved~(E02, Q04, Q05, E08, E11). The participants also expressed various ideas to improve social media in terms of planning tools such as an integrated planning module~(E07, E08) with automated scheduling~(E12). Lastly, tools for the creation of content were requested~(Q01, Q03, E07, E08, E12), comprising integrated and/or automated text, image, and/or video editing~(Q01, Q03, E07, E12) to streamline collaborative tasks constituting activist work. These recommendations have already been adopted by external services and, to some extent, even by social media platforms themselves.

\added{\textbf{Increasing Customizability:}}
Activists requested enhanced customizability, especially for display options and screen-reading settings~(Q01, Q04, Q05, E12). Activists considered customizable display options important for allowing audiences to tailor the visual layout to their preferences. This flexibility would help activists reduce the impact of conflicting content display needs from their audience. Activists also emphasized the need for customizability for their activism-related content consumption~(Q01, Q04, Q05, E12). For example, one neurodiverse activist emphasized the need to adjust the physical distance between posts to avoid feeling overwhelmed by the content feed.

\section{Discussion}
Investigating activists' use of social media is crucial for understanding how these platforms can support or hinder social movements. While issues like hate speech and unequal platform visibility have been explored~\cite{Castillo-Esparcia2023Evolution, Momeni2017Social, Sannon2023Disability}, less attention has been given to activists' visions for addressing these challenges. 

This paper fills that gap by examining the \added{needs} and requested features of activists from two social movements that differ in size, predominant protest form, and their social media strategies. Our key findings are that activists high-level \added{needs} are \textit{On- and Offline Safety}, \textit{Enhanced Content Presentation}, and \textit{Optimized Workflows}, and that activists request specific features to \added{meet these needs}. 

In the discussion, we will contextualize our findings within the existing literature. We will compare the answers of the environmental movement with those of the queer-feminist movements and offer explanations for the observed differences. Finally, we present strategies to \added{address the needs} of the activists. To this end, we offer design, research \added{and strategy} recommendations. 

\subsection{Comparison of \added{Needs} and Feature Requests to Previous Studies}

\subsubsection{On- and Offline Safety}
Participants requested changes to social media platforms to increase on- and offline safety. 

\added{\textbf{Physical Safety:}}
Suggested features relating to increased \added{physical} safety primarily surround revised approaches to anonymity. Activists deemed features such as non-public comments, self-deleting messages, and the option to choose from several levels of anonymity an important measure to ensure safety. These requested features predominantly enhance their control over how and which information can be tied to which aspect of their identity. The right to control personal information is defined as privacy~\cite{moore2008defining}. Activists thus want to increase privacy options to reach anonymity when deemed necessary. \added{While anonymity can protect activists, particularly those operating in politically repressive contexts, it can simultaneously be harmful when misused by others to perpetrate online assaults. For instance, the same anonymity that shields activists from retaliation may also embolden individuals to engage in harmful behavior without accountability~\cite{Barlett2018Social, dwivedi2018social}.} 

\added{This tension has been identified by} the activists. \added{Some assessed }anonymity as a key factor for enabling online assaults. Consequently, several activists requested that social media accounts should be tied to the personal identity of the user, e.g. through disclosing parts of their identity. Further, the use of multiple accounts to evade consequences by violating platform guidelines has been criticized. \added{The lack of anonymity is a double-edged sword for activists: While a high-profile presence can mean protection due to increased public interest, there is also an increased risk of being exposed to hate speech, attacks, censorship or (governmental) repression. For instance, linking social media accounts to user identities was proposed as a measure to combat abuse, but could expose activists to greater risks in contexts of censorship or political repression. In such environments, losing anonymity might equate to losing their platform. Similarly, suggestions to discourage the use of multiple accounts for accountability purposes could unintentionally restrict activists' ability to circumvent restrictions or evade surveillance. These tensions reflect broader dilemmas surrounding anonymity policies. In addition to that, implementing such features may conflict with social media platform's business models. The request of some activists for identity verification could lead to user backlash or reduced engagement, particularly in contexts where anonymity is crucial for safety. Platforms must weigh these trade-offs carefully, as stricter identity requirements may align with regulatory demands or brand safety concerns but risk diminishing user engagement. A recent example of the double-edged nature of changes in anonymity policies is the modification of the ban function on X. For public accounts on X, users who are blocked can still access content but cannot engage with it. On the one hand, this seems to directly fit the requests of our participants, as Q01 said: ``I don't want the hate, but I want them to know about it.'' On the other hand, this change in functionality can be criticized for empowering bad actors, such as stalkers or harassers by still allowing them to view the content of the respective account. Users can still publicly react to content by uploading screenshots of the respective post and comment on it.}

Previous research on anonymity in social media primarily examined its link to incivility~\cite{andalibi2018anon,  Barlett2018Social, Jaidka2021Beyond, warren2018youngadultsanonymity}. A survey study found that young adults generally believe social media is better (relating to cyberbullying) when real names are used ~\cite{warren2018youngadultsanonymity}. This aligns with findings that link user anonymity to negative behaviors such as cyberbullying~\cite{Barlett2018Social}. However, other studies have found no significant relationship between anonymity and incivility on social media~\cite{andalibi2018anon, Jaidka2021Beyond}. This shows that the conflicting perspectives of our activists are echoed in previous research. Interestingly, research has shown that combining personal anonymity with social identifiability can reduce incivility, as demonstrated in discussions about gun rights among a U.S. sample~\cite{Jaidka2021Beyond}. This concept of selective identity disclosure, or ``meronymity'', has been explored as a potential strategy to balance anonymity and accountability~\cite{soliman2024mitigating}. Furthermore, prior studies have advocated for a more comprehensive approach to online safety, proposing frameworks that address digital security, digital privacy, offline safety, and community standards~\cite{Redmiles2019I}. This approach likely addresses the needs of our participants more holistically. \added{However, platforms may be hesitant to adopt such comprehensive frameworks due to the significant resources required to support enhanced safety features.} Our study extends \added{previous} insights by documenting the unique perspective of activists navigating the complex tension between anonymity and accountability in their (online) activist work.

\added{\textbf{Reducing Exposure to Hate Speech: }}To reduce the exposure to hate speech the activists requested reliable filters for hate speech, educational efforts, and increased consequences for users spreading hate speech. Participants' requests align with the research thread on mitigating hate speech centered around hate speech detection~\cite{Rathod2023From, Frenda2019Online, Rodríguez‐Sánchez2020Automatic}, understanding the dynamics of its spread~\cite{maarouf2024hate, solovev2022hate}, and legal efforts~\cite{Fabbri2023DSA}. While automated detection systems have achieved higher accuracy, they still face significant challenges, including poor generalization, bias against specific communities, and the ongoing debate over free speech~\cite{Poletto2020Resources}. Previous research on the impact of hate speech filtering combined with explanations for content removal, as suggested by our participants, hints towards positive effects on hate speech both for the respective users as well as bystanders~\cite{Jhaver2024bystander}. However, the implementation of such measures must be critically evaluated. The effectiveness of automated filters can vary significantly across different contexts, and there is a risk that overly broad or poorly designed filters could inadvertently suppress legitimate discourse or disproportionately impact marginalized communities. \added{Other activists have reported their perception that under the guise of content governance, their political stance has been censored~\cite{abokhodair2024opaque, elmimouni2024shielding}. Research suggests, that there are political biases in content moderation, which is of particular importance to activists~\cite{abokhodair2024opaque, elmimouni2024shielding}. Elmimouni et al. therefore argue strongly for greater transparency in content moderation. Social media platforms may view both advancing hate speech filtering and increased transparency in content moderation as resource-intensive. Improving hate speech detection requires robust AI models trained on diverse datasets, which demand substantial computational resources, continuous refinement, and human oversight. And even with these efforts, hate speech moderation remains an issue that can only be addressed to some extent, as biases in training data, context sensitivity, and evolving language patterns pose persistent challenges. Moreover, platform-wide moderation policies have to navigate a complex landscape of highly diverse political contexts and cultural and societal norms. Thus, such policies inevitably remain limited. Though better hate speech handling and increased transparency could improve user retention in safer spaces, it may also conflict with engagement-driven business models.} Thus, while our participants' requests are grounded in existing research, careful consideration is needed to balance the enforcement of hate speech policies with the protection of free expression. 

Activists perceive on- and offline safety to be highly interconnected, view social media as a key factor in their on- and offline safety, and emphasize a nuanced balance between anonymity and accountability. This aligns with previous research on the role of anonymity in hate speech and online incivility and extends these approaches by emphasizing the interconnectedness of on- and offline safety for activists. \added{This also reveals critical tensions within the operational and financial priorities of the platforms, requiring careful consideration of both feasibility and impact.}

\subsubsection{Enhanced Content Presentation}
\added{Participants requested changes to social media platforms to allow for enhanced content presentation.}

\added{\textbf{Reaching Target Audience:}} Activists reported the importance of reaching their audience. \added{While reaching broader audiences, including bystanders, is essential for activist efforts \cite{Siddarth2020Engaging}, it inevitably exposes the content to a larger number of dissenting viewers. This likely further increases the possibility for the activists to become targets of hate speech, which creates tension with the need to reduce exposure to hate speech. Hate speech, as previously defined~\cite{zhang2019hate}, is clearly distinct from criticism, and most social media platforms have regulatory policies against hate speech in place~\cite{meta_hatespeech, tiktok_guidelines, x_rules}. Activists must accept that greater visibility inevitably leads to more criticism. However, the need for effective protection against hate speech remains crucial and should not conflict with the aim for broader reach. But, the activists reported}  lacking algorithmic visibility, which is a documented issue\added{~\cite{Castillo-Esparcia2023Evolution, elmimouni2024shielding}}. 
The requested changes mainly relate to increasing transparency and agency in the interaction with the recommendation system. Previous research emphasized the positive effect of increased transparency of recommendation systems of social media, such as increased awareness of how the system works, and increased trust and perceived usefulness~\cite{Rader2018Algo, Shin2020Algo}. Nevertheless, it has also been critically discussed in terms of its limitations, such as balancing cognitive overload and transparency, its technical limitations, and its inability to act as a sole mechanism to ensure accountability~\cite{Ananny2018Limits}. Lack of transparency has also been assessed as necessary \added{from the platform's perspective} to protect intellectual property and to prevent users from gaming the system~\cite{jhaver2018algorithmic}. \added{Thus, it is likely not in the interest of social media platforms to increase transparency relating to their recommendation system as it has inherent technical limitations, requires additional resources, constitutes a competitive disadvantage, and enables users to game the system~\cite{Ananny2018Limits}.}

The request for increased agency over the recommendation system on social media has also been scientifically investigated~\cite{kitchin2019thinking, van2018platform}. Through recommendation systems, users' agency is diminished~\cite{van2018platform}. However, through manual personalization and implicit actions such content engagement feedback loops emerge~\cite{kitchin2019thinking}, offering some level of agency over the displayed content. However, the participants perceived these possibilities as not providing sufficient agency. Activists expressed concerns about their limited agency within algorithmic systems, something not emphasized in prior studies. Their demands for more robust agency highlight a desire for functional autonomy that extends beyond mere personalization. \added{Social media platforms optimize content recommendations to maximize user engagement and time spent on the platform. Increased user control over recommendations could conflict with these goals, potentially reducing platform usage~\cite{zhang_monitoring_2022} in addition to posing technical implementation challenges.}
The Digital Services Act~(DSA) of the European Union sets transparency and auditing requirements for recommendation systems, allowing users to modify the parameters that influence the content they see. Fabbri~\cite{Fabbri2023DSA} claims, that to make the DSA’s provisions effective, policymakers should ensure users are provided with clear explanations of how recommendations work and enable them to directly adjust the strategies that guide these systems. This directly aligns with the participants' demands for greater agency and transparency.

\added{\textbf{Controlling Scrolling Behavior:}} Previous research supports \added{our findings} by showing the link \added{of the infinite scroll} to addiction, dissociative states~\cite{Lyons2022Design}, and feeling trapped and regretful~\cite{Rixen2023The}. Some activists requested the option to switch off the infinite scroll manually, and the introduction of timed digital nudges that prompt users to take a break after a specific duration of scrolling. Scholars have criticized this \added{kind} of interventions for not addressing the features producing the unwanted behavior in the first place~\cite{zhang_monitoring_2022}. They \added{suggest} to change the internal mechanisms producing the unwanted behavior over external mechanisms. \added{However, infinite scroll is associated with longer user engagement~\cite{Rixen2023The}, aligning with the business models of the platforms. Altering or disabling this feature may reduce the time users spend on the platform, potentially making such changes less attractive to platform stakeholders. While implementing changes to infinite scroll may conflict with platform business models, we argue that addressing the root causes of problematic usage is essential to fostering healthier interactions, especially for activists whose work relies on focused and intentional social media use.}  

\added{\textbf{Managing Misinformation:}} To better manage misinformation, participants requested better regulation. The requested features comprise better filtering and better labeling of misinformation. Previous research has already widely addressed options for automated misinformation detection~\cite{Bagozzi2024mis, Islam2020Deep}, investigated the impact of labeling of misinformation~\cite{barman2024misinf, Guo2023mis, Roozenbeek2023Countering, Yaqub2024Mis}, and compared the effectiveness of various kinds of labeling and other platform inventions~\cite{jia2022Understanding, Guo2023mis}. Furthermore, Instagram, the context for which labels were requested, already labels misinformation. This could indicate that participants did not yet see labeled misinformation or that they still frequently encounter misinformation without labels, hinting toward insufficient labeling. \added{However, from the platform perspective this presents both advantages and challenges. On the one hand, rigorous removal of misinformation might increase brand safety. On the other hand, misinformation detection is a resource-intensive endeavor, and platforms may weigh these investments against their impact on user engagement, particularly if labels deter interaction with certain types of content~\cite{Roozenbeek2023Countering}.}
Interestingly, the participants did not wish for features enhancing their options to address misinformation or other forms of crowd-sourced fact-checking, which has been shown to yield positive results~\cite{Heuer2024Re, jahanbakhsh2023misinf, Jahanbakhsh2024A, Martel2023Crowds, Pinto2019Towards}. 

Overall, activists’ demands align with existing research but emphasize persistent gaps in platform accountability, transparency, and agency, especially relating to greater control over algorithmic systems and better management of misinformation. \added{These demands highlight disparities between the activists' needs for empowerment and the platforms' existing priorities, focusing on user engagement, and operational efficiency.}

\subsubsection{Optimized Workflows}
\added{Activists highlighted their need for optimized workflows, consisting of the need to reduce workload and to increase customizability.}

\added{\textbf{Reducing Workload:}} Nearly all participants mentioned the challenge of limited time resources, making workload reduction a crucial \added{need}. Previous research on activist workload on social media has generally emphasized the advantages of these platforms, such as their role in facilitating political activism~\cite{Keith2023The, Momeni2017Social} and enhancing the effectiveness of activist efforts through strategic use~\cite{Bisafar2020act}. The high workload of activists has been acknowledged in the literature, with studies highlighting behind-the-scenes work~\cite{Callaghan2011The} and the risks of unequal distribution of workload~\cite{Lauby2023GENDER}. This high workload has rarely been linked to social media use \cite{Larsen_Ledet2023}. Our findings add a novel dimension to this discourse by revealing that activists perceive a significant portion of their workload as directly tied to social media and they are calling for concrete changes to alleviate this burden. These insights underscore the need to consider the practical needs of activists who rely on these tools for their work. \added{However, activists' needs for workload reduction may differ from those of other user groups, requiring platforms to balance diverse priorities. Implementing activist-specific tools could demand significant resources, that platforms may hesitate to allocate given their focus on broader user engagement and their current policy not to recommend political content~\cite{instagram_political_content}.}

\added{\textbf{Increasing Customizability:}} Lacking customizability is also one tangible challenge for activist workflows. Customizability is a frequently documented wish of various user groups on social media~\cite{Armstrong2024custom, Lim2023TBI, pena2023autism, Sunkara2023Custom}. For activists, customizability impacts their ability to efficiently manage their work and ensures that their content is accessible to a diverse audience. Our findings show the alignment between activists' demands and broader research findings, highlighting the critical need for more flexible and personalized interfaces. The inability to personalize the interface ---whether through modifying how content is displayed, customizing notification settings, or managing screen-reading options --- limits their ability to target and engage audiences in a way that fits their unique workflow requirements.
\added{Although customizability is a shared demand across various user groups~\cite{Armstrong2024custom, Lim2023TBI, pena2023autism, Sunkara2023Custom}, the specific features and the way in which they require customization may vary greatly.  Meeting these diverse needs is resource-intensive because it involves developing, testing, and maintaining a range of flexible features that cater to different preferences. This can significantly increase platform complexity and operational costs, which may not generate the return on investment necessary from the platform's perspective.}

\added{Overall, our findings highlight activists' unique challenges with social media workflows, particularly the high workload and limited customizability. Our findings extend existing research by demonstrating that activists report that using social media for activism constitutes a significant workload due to platform inefficiencies. This underlines the need for targeted solutions. These insights underscore the tension between activists' specific needs and platforms' broader priorities.}

\subsection{Comparison of Queer-feminist Movement and Environmental Movement}
This section highlights key differences between the environmental and queer-feminist movements in social media use. Differences in three central dimensions were apparent (an overview can be found in Table \ref{table3}).

\begin{table*}[ht]
\caption{The differences between the activists from the environmental movement and the queer-feminist movement in three central dimensions were identified through semi-structured interviews with a card-sorting task with 14 activists.
}
\Description{This table has three columns: Dimensions, Environmental Movement, and Queer-Feminist Movement. The Dimensions column lists 3 items: Physical Safety; Avoiding Exposure to Hate Speech and Reducing Workload. The Environmental Movement Column lists three items: Less direct personal threats; Focus on existing moderation features; Focus on planning-related tools. The Queer-Feminist Movement column has three items: All activists report in-person assaults; Emphasis on better platform regulation; No focus and planning-related tools}
    \begin{tabular}{l l l}
    \toprule
    \textbf{Dimensions} & \textbf{Environmental Movement} & \textbf{Queer-Feminist Movement} \\
    \midrule
    \textbf{Physical Safety} &	Fewer direct personal threats &	All activists report in-person assaults \\
    \textbf{\added{Reducing} Exposure to Hate Speech} &	Existing moderation features &	Emphasis on better platform regulation\\
    \textbf{Reducing Workload} &	Focus on planning-related tools  &	No focus on planning-related tools \\
    \bottomrule
    \end{tabular}
    \label{table3}
\end{table*}
 
\subsubsection{\added{Physical Safety}}
The most striking difference becomes apparent regarding the \added{need for} physical safety. Here, all participants from the queer-feminist movement report that they lack physical safety relating to their activism. One contributing factor is that activists from the queer-feminist movement either were personally confronted with or witnessed in-person assaults. Predominantly activists from the queer-feminist movement perceived the combination of both being present on social media and performing activist actions mostly in person as unsafe. One explanation could be the most dominant protest form of these movements: While the environmental movement mostly mobilizes for large demonstrations, the queer-feminist movement mostly performs activist actions in small groups at locations at which street harassment took place. Frequently, incidents of street harassment take place at the same location, putting the activists in more vulnerable situations. Another potential explanation could be the difference in the societal perception of the relevance of their causes. The goal of combating climate change is potentially more appealing to the majority. In Germany, the Green Party has been advocating for climate protection for over four decades, and the Paris Agreement underscores the international commitment to addressing climate change. While adherence to such agreements may vary, climate change holds substantial social relevance, at least within public discourse. In contrast, queer-feminist activism, although growing in visibility, remains more marginalized. Movements like \#MeToo demonstrate how issues related to gender-based violence and systemic sexism continue to be sidelined. The consequent removal of identifiable content on social media is also reported by four of the five activists from the queer-feminist movement, while only by one activist from the environmental movement.
\subsubsection{\added{Reducing Exposure to Hate Speech}}
The two movements also differed in terms of feature requests addressing hate speech. While the members of the environmental movement focused more on existing features for moderation, the members of the queer-feminist movement emphasized the need for better platform regulation. One potential explanation could be that the queer-feminist movement is more exposed to hate speech. Another explanation could be, that because they already regularly receive reports of hate speech through their form of activism, they want to avoid any additional encountering of hate speech.

\subsubsection{\added{Reducing Workload}}
Requested features to reduce workload also differed. Features requested by members of the environmental movement focused on improvements in planning-related activities. \added{Potentially, because} the environmental movement frequently operates on national and international levels. The queer-feminist activists did not ask for features to facilitate content production on social media. This is surprising since they frequently use external services for content production. This absence is noteworthy and might suggest that while they engage heavily in content creation, they are already somewhat content with their workflow relating to content production, potentially because external services for image and video editing are employed.

Thus, differences between the two social movements are apparent, particularly relating to their concerns about physical safety. Understanding the differences between activist groups can guide the design of tools that are tailored to their specific needs, enhancing safety, reducing exposure to hate speech, and improving workflow efficiency. This highlights the importance of addressing the diverse realities of different movements.

\subsection{Design, Research \added{and Strategy} Recommendations}
The following section outlines design and research recommendations to address the \added{needs} identified in this research. These recommendations are categorized according to the high-level \added{needs} on- and offline safety, enhanced content presentation\added{,} and optimized workflows. These recommendations are informed by participant-requested features, supported by previous literature, and, in some cases, extend beyond these foundations to suggest new avenues for addressing the identified challenges.

\subsubsection{On- and Offline Safety}
The activists highlighted the need for physical safety and a reduced exposure to hate speech as particularly important.

\added{\textbf{Physical Safety:}}
Partly contradicting features relating to anonymity were requested. A possible design recommendation addressing these conflicting demands around anonymity is to implement adaptive anonymity settings. These settings could allow users to choose different levels of anonymity based on the context of their interactions, such as full anonymity in public forums and varying levels of identifiability in different groups. Such adaptive anonymity settings could be informed by insights from previous research on meronymity\added{~\cite{Jaidka2021Beyond, soliman2024mitigating}}. \added{However, the potential of implementing adaptive anonymity settings raises questions about the viability of such features. Features like adaptive anonymity may face challenges in adoption because they require significant resources, including developers and computational capacity for dynamic anonymity management. To encourage adoption, future research could explore how such safety features could be framed to align with stakeholder goals, such as enhancing user trust or compliance with regulatory demands. 
An approach for activists to navigate these constraints could be to leverage decentralized platforms or adopt encrypted communication tools. Activists reported that they felt forced to be anonymous for safety reasons but wished to produce more personal content conveying their stance. Working with avatars or other forms of personal but unidentifiable content could prove helpful in this endeavor and is a well-established activist practice~\cite{gerbaudo2015protest}.
Collaborative frameworks between researchers and activists could facilitate the co-creation of features that align with safety needs. Such co-creation approaches have already proven successful in previous work~\cite{molina2022cocreating}, also relating privacy and security of activists~\cite{tadic2023design}}.

These approaches could help balance the need for safety with the protective benefits of anonymity, enabling activists to engage in activism more safely. \added{Future research should explore how to best balance safety, accountability, and freedom of speech within adaptive anonymity settings. This could involve exploring various levels of transparency and control, allowing activists to adjust their anonymity settings in response to specific audiences or objectives.}

\added{\textbf{Reducing Exposure to Hate Speech:}} Social media platforms could provide more extensive educational explanations when content is removed, helping users understand the reasons behind these actions and preventing future misconduct. Another approach could be to nudge users towards civil behavior, as tested in previous work~\cite{bossens2021improving, Seering2019Ensuring}. Through priming positive emotions or constructive discourse impulses, civil behavior and constructive discourse could be fostered~\cite{bossens2021improving, Seering2019Ensuring}, which is of particular interest to activists. Also, personalized filtering of hate speech using AI approaches could prove efficient in addressing the needs of the activists. Through a user-friendly dashboard, activists could set preferences for filtering specific types of hate speech, with options for sensitivity levels ranging from high to low. The system would use real-time AI algorithms to \added{assess} content, learn from user feedback, and adapt its filtering over time. \added{It should be noted however, that current ML detection systems still suffer from significant performance issues such as overmoderating legitimate discourse as well as undermoderating hate speech that expresses itself more implicitly. Examples for implicit hate speech are coded or indirect language like stereotypes, sarcasm, irony, humor, or metaphors~\cite{Elsherief2021Latent, Jafari2023Fine-Grained}. Context is another important aspect that has been shown to significantly improve hate speech detection and annotation accuracy~\cite{Ljubesic2022Quantifying, Yu2022Hate}. Allowing users to manage hate speech handling across self-defined contexts presents an opportunity to enhance moderation. By allowing users to individually adjust moderation mechanisms for their feeds, specific accounts, and comments on particular posts, users can more effectively integrate contextual considerations into moderation efforts. This approach could effectively bridge the concepts of personal account moderation and personal content moderation, as discussed in \cite{Jhaver2023}.}
\added{While personalized filters offer substantial potential, their realization requires a robust ecosystem of technical and social support. Implementing such tools would necessitate integrating real-time AI processing pipelines, secure data handling for user-specific preferences, and mechanisms for user feedback to refine algorithms continuously. Collaborative frameworks between platforms, researchers, and civil society groups could help ensure that these tools are designed and deployed effectively while minimizing unintended harms.}

Thus, \added{changes in activist strategies}, the implementation and continuous evaluation of adaptive anonymity settings, and personalized hate speech filters are promising directions to enhance activists' online and offline safety.

\subsubsection{Enhanced Content Presentation}
Enhanced content presentation is crucial for the activists. Specifically, activists requested more transparency and agency relating to the recommendation system. 

\added{\textbf{Increased Transparency:}} Activists requested insights into the algorithmic curation system. Platforms could offer user-friendly explanations of how algorithms influence content presentation. There exists, however, extensive research on how to best implement these explanations, highlighting the complexity of the issue~\cite{garfinkel2017toward, Rader2018Algo, watson2019addressing}. From the activists' perspective, simplified insights into why posts gain or lose visibility would be of particular interest. Transparency dashboards could offer a clear overview of how and why specific content performs in certain ways. From the platform’s perspective, offering transparency into algorithmic systems presents several challenges. While simplified insights may be helpful for activists, platforms must balance providing transparency with protecting proprietary algorithms and avoiding potential exploitation of the system. 

\added{\textbf{Increased Control:}} Activists further requested more control over the algorithmic curation system of social media platforms. A notable request was to return to a chronological feed, which some participants viewed as something they had more control over. To this end, social media platforms could implement options to toggle between algorithmic and chronological feeds \added{(as for example already implemented by Instagram)}. This would, however, not ensure that their target audience will use it in the same way. Recent research~\cite{Guess2023Fb} found that switching to a chronological feed increased exposure to political and untrustworthy content while reducing uncivil content, particularly on Facebook. Despite these changes, key attitudes like polarization and political knowledge remained largely unaffected over the 3-month period~\cite{Guess2023Fb}. This suggests that while the chronological feed may offer a more predictable content flow both for them and their target audience, it may not necessarily drive substantial shifts in user attitudes or deeper political engagement. This points to a potential gap between activists' expectations and the actual impact of such a change. While decentralized platforms with chronological feeds exist, they are underused by activists. This is probably because they have a smaller reach and engagement than mainstream platforms. To address these concerns, activists could consider expanding their presence across multiple platforms, including those with chronological feeds, and synchronizing their content across these networks. This strategy could help mitigate the perceived drawbacks of mainstream platforms.
Further, tools that help activists measure and track the impact of their content beyond simplistic engagement metrics could increase the agency of the activists. This could include user-friendly dashboards that help activists analyze content performance across different channels or tools that offer real-time feedback on how to adjust content to improve visibility. The platforms' business model is fundamentally focusing on user attention, and recommendation systems are optimized to prioritize content that keeps users engaged for longer periods. This often limits the reach of activist content, which may not be as algorithmically engaging as entertainment or commercial posts. Platforms are designed to keep users on-site, meaning activist content that does not align with these objectives is less likely to receive a favorable placement. Given this dynamic, one could explore more strategic solutions that work within these platform constraints. For example, research could focus on how activists might leverage collaborative content creation or cross-platform promotion to boost visibility.  

Additionally, participants raised concerns about the algorithmic reduction of political content on Instagram. Activists can counteract this by disabling such mechanisms in the settings section. By encouraging their followers to do the same and explaining to them how to do so they could help restore the visibility of political content.

Thus, advancing research on transparency and agency in algorithmic curation is promising for enhancing the content presentation of activists. Tools that track content performance beyond basic metrics and the expansion across multiple platforms using collaborative strategies could help overcome visibility challenges. Activists can also address the reduction of political content by adjusting platform settings and guiding their audience to do the same.

\subsubsection{Optimized Workflows}
Activists emphasized the need to reduce workload and increase customizability.

\added{\textbf{Reducing Workload:}} Our findings suggest that improvements in content production and internal coordination are particularly crucial \added{for activists. However,} implementing workload-reduction tools tailored to activist use cases may be challenging for platforms due to the broad spectrum of user needs.  \added{But many external services offer support with both content production (e.g. Canva~\cite{canva}, PicMonkey~\cite{picmonkey}, Unsplash~\cite{unsplash} for images and graphics, and Animoto~\cite{animoto} or Biteable~\cite{biteable} for videos) and cross-platform scheduling and management (e.g. Buffer~\cite{buffer}, CoSchedule~\cite{coschedule}, or Sprinklr~\cite{sprinklr}). Buffer, for example, offers the possibility of publishing and planning content for several social media platforms through the same dashboard and allows for collaboration. These aspects have been emphasized by our participants as particularly important and could thus prove helpful in reducing the workload for activists. 
Despite their potential utility, these tools are not widely adopted by activists. Possible reasons include the brand-oriented marketing approach of these services, which may not resonate with activists who view their mission as separate from conventional branding and marketing. Additionally, while some tools offer free basic versions, essential features often require paid upgrades, which could be a barrier for activists. Adopting these tools also demands an initial time investment to integrate them into existing workflows. While potentially saving time in the long run, this initial time investment might deter activists from deploying such services. As the activists collaborate with many other local chapters, establishing the use of new tools within a larger network might be particularly challenging. Investigating why activists are not utilizing these existing tools, despite their apparent need, should be a focus of future research.} Co-designing tools with activists to directly address workload reduction could \added{also prove to} be highly beneficial. 

\added{\textbf{Increasing Customizability:}} Enhancing customizability could allow activists to manage content more efficiently and focus on their core activities. Future research should explore the specific customizability needs of different activist movements to better understand how these features can be tailored. Enhancing customizability could greatly benefit activists, enabling them to manage their workflows more efficiently. Future work should focus on understanding \added{the} specific \added{customizability} needs. \added{Co-designing tools with activists to gain insights into the specific customizability needs of activists and how they relate to previously identified customizability of other user groups could prove highly beneficial. }

\subsection{Limitations \& Future Work}
The primary objective of this study was to investigate the \added{needs} and feature requests of different social movements \added{for their social media activism}. \added{
By focusing on two movements with distinct characteristics, this research provides foundational insights into the diverse ways activists engage with social media.}
Participants were recruited from local chapters of social movements based in Germany, which might not reflect the experiences of activists in other cultural or geopolitical contexts. Additionally, our sample is WEIRD—Western, Educated, Industrialized, Rich, and Democratic—potentially limiting the generalizability of our findings. Another limitation is that participants predominantly \added{used} Instagram \added{for their activism}, which may not capture the varied experiences of activists using other platforms. \added{We only included responses related to activist use of social media in the analysis. However, our participants may not always have been able to consistently distinguish between activist and recreational social media use.} 

\added{The selection of the environmental and queer-feminist movement further provided an opportunity to explore various activists’ needs. While our findings highlight valuable insights, they leave open questions about the relative influence of practical differences, such as movement size or social visibility, versus the intrinsic nature of the movements themselves. Addressing these questions in future work could extend the groundwork laid by this study, enriching the understanding of how diverse social movements navigate and shape their digital environments.}

Future research should include activists from a wider range of socio-economic, cultural, and geopolitical contexts, using different social media platforms \added{and with varying thematic focuses or organizational structures}. Mixed-method approaches, such as qualitative interviews combined with quantitative surveys, could be particularly useful. While our qualitative interviews provide deep, context-rich insights, \added{surveys could assess the broader relevance of the requested features and help prioritize needs}. Expanding the scope in this way would enhance the applicability of the results and offer a more comprehensive understanding of the features activists need.

\added{Future work could incorporate participant observation or field studies during activist actions to capture real-time social media use and contextual factors. This approach would complement interview insights and support a more grounded understanding of needs. Co-design methods, where activists collaborate in developing tools and platforms \added{addressing the identified needs}, could further align solutions while addressing platform constraints. By bridging the gap between activist needs and platform capabilities and priorities, co-designing third-party services may also create practical and impactful interventions within the broader digital landscape.}

\section{Conclusions}
In this study, we explored the perspectives of activists from an environmental movement and a queer-feminist movement through semi-structured interviews, uncovering their \added{needs} for and desired features of a future social media platform. Our findings reveal that activists' high-level \added{needs} are on- and offline safety, enhanced content presentation, and optimized workflows. These are socio-technical challenges, where activists can be supported through further research and improvements to social media platforms.

Our research contributes to a deeper understanding of the relationship between activism and social media, providing concrete recommendations for platform design and future research. By emphasizing the critical need for both on- and offline safety, this study paves the way for developing safer social media environments for activists. These insights can inform future platform innovations, helping to ensure that social media supports—not hinders—the important work of social movements in driving meaningful change.
\appendix
\added{
\section{Appendix}
\subsection{Activist Tasks}}
\label{ActivistTasks}
\added{
We analyzed the tasks that the activists report as constitutive of their activist work and their description of how they used Instagram and Messenger Services to fulfill these tasks. We categorized the tasks into general categories of activist work. The features of Instagram and the messenger services (WhatsApp, Telegram, and Signal) are explicitly mentioned by the participants. We categorized the features of Instagram into Content Creation (consisting of posts, stories, reels, live streaming, and editing tools), Engagement \& Interaction (consisting of likes, comments, shares, polls/quizzes, DMs, and tagging), Discovery \& Browsing (explore tab, hashtags, location tags, search functionality, and suggested posts), Account Management (consisting of profile customization options, private/public accounts, close friends list, account switching, and insights) and Content Moderation (consisting of mute, block, restrict, report, privacy settings, archive). The features of messenger services were categorized into Individual Chats (consisting of test/voice messages, voice/video calls, and file sharing), Group Chats (consisting of test/voice messages, voice/video calls, and file sharing), Channels (consisting of text messages and image/video/location sharing), Privacy \& Security (consisting of self-deleting messages, usernames and chat lock) and Collaborative Tools (consisting of Polls and Quizzes).
}

\begin{acks}
We would like to thank all participants for sharing their perspectives with us. We would like to thank our anonymous reviewers for their very helpful and constructive feedback. This work is funded by the German Federal Ministry of Education and Research and the European Union - NextGenerationEU.
\end{acks}

\bibliographystyle{ACM-Reference-Format}
\bibliography{references}

\end{document}